\documentclass[final]{elsart3p}

\usepackage{graphicx}

 \journal{Comp. Mater. Sci.}
\usepackage{dcolumn}
\usepackage{bm}

\begin{document}

\begin{frontmatter}

\title{Half-metallic ferrimagnetism in the [Sc$_{1-x}$V$_x$]C and [Sc$_{1-x}$V$_x$]Si alloys
adopting the zinc-blende and wurtzite structures from
first-principles}

\author[Gebze]{K. \"Ozdo\~gan}\ead{kozdogan@gyte.edu.tr}
\author[Julich,Fatih]{E. \c Sa\c s\i o\~glu}\ead{e.sasioglu@fz-juelich.de}
\author[Patras]{I. Galanakis\corauthref{cor}}\ead{galanakis@upatras.gr}

\address[Gebze]{Department of Physics, Gebze Institute of Technology,
Gebze, 41400, Kocaeli, Turkey}
\address[Julich]{Institut f\"ur Festk\"orperforschung, Forschungszentrum
J\"ulich, D-52425 J\"ulich, Germany}
\address[Fatih]{Fatih University,
Physics Department, 34500, B\" uy\" uk\c cekmece,  \.{I}stanbul,
Turkey}
\address[Patras]{Department of Materials Science, School of Natural
  Sciences, University of Patras,  GR-26504 Patra, Greece}
  \corauth[cor]{Corresponding author. Phone +30-2610-969925,
Fax +30-2610-969368}

\begin{abstract}
Employing first-principles calculations we study the structural,
electronic and magnetic properties of the [Sc$_{1-x}$V$_x$]C and
[Sc$_{1-x}$V$_x$]Si alloys. In their equilibrium rocksalt
structure all alloys are non-magnetic. The zincblende and wurtzite
structures are degenerated with respect to the total energy. For
all concentrations the alloys in these lattice structures are
half-metallic with the gap located in the spin-down band.  The
total spin moment follows the Slater-Pauling behavior varying
linearly between the -1 $\mu_B$ of the perfect ScC and ScSi alloys
and the +1 $\mu_B$ of the perfect VC and VSi alloys. For the
intermediate concentrations V and Sc atoms have antiparallel spin
magnetic moments and the compounds are half-metallic ferrimagnets.
At the critical concentration, both [Sc$_{0.5}$V$_{0.5}$]C and
[Sc$_{0.5}$V$_{0.5}$]Si alloys present zero total spin-magnetic
moment but the C-based alloy shows a semiconducting behavior
contrary to the Si-based alloys which is a half-metallic
antiferromagnet.
\end{abstract}

\end{frontmatter}

\section{Introduction}\label{sec1}

Half-metallic ferromagnets have attracted considerable attention
during the last decade due to their potential applications in
magnetoelectronic devices \cite{Zutic}. The term ``half-metal''
was initially introduced by de Groot and collaborators in 1983 to
denote the peculiar behavior exhibited by a Heusler compounds:
NiMnSb \cite{deGroot}. The have found using first-principles
calculations that the majority-spin band was metallic while the
minority-spin band was semiconducting leading to 100\%\
spin-polarization of the electrons at the Fermi level. This
behavior was later on confirmed both by infrared absorption
\cite{Kirillova} and by spin-polarized positron annihilation
\cite{Hanssen} experiments.

Although Heusler alloys have attracted a lot of interest as
potential half-metallic systems, the discovery of Akinaga and his
collaborators in 2000 has shown the possibility to grow new
half-metallic systems in metastable structures usually adopted by
thin films \cite{Akinaga}. They have shown that the CrAs/GaAs
multilayers are ferromagnets and surprisingly the X-ray
diffraction measurements suggest that CrAs adopts the lattice
structure of GaAs and grows in the metastable zincblende
structure. Moreover SQUID measurements have shown that CrAs
exhibits an integer total spin magnetic moment of 3 $\mu_B$ per
unit cell \cite{Akinaga}. These findings have intensified the
interest on transition-metal pnictides and chalcogenides like CrAs
and CrSe which crystallize either in the zinc-blende or wurtzite
structures of binary semiconductors and an extended review can be
found in reference \cite{Mavropoulos}. Galanakis and Mavropoulos
have studied using first-principles calculations several such
compounds and have determined the lattice constants for which
half-metallicity is present \cite{GalaZB}. Moreover they have
explained the gap in terms of the $p-d$ repulsion; the $p$
orbitals of the $sp$ atom hybridize with the $t_{2g}$ orbitals of
the transition metal atoms creating three bonding and three
antibonding states. The gap is created between these states. The
$e_g$ orbitals of the transition metal atom are very localized in
energy since they do not hybridize with other orbitals and they
are placed above the Fermi level in the spin-down band. Their
relative position with respect to the antibonding $p$-$t_{2g}$
orbitals depends on each system. It has been also shown in the
same reference that except the three bonding $p$-$t_{2g}$ bands
below the Fermi level exists also a deep $s$ band. Thus, since
there are exactly four spin-down occupied bands, the total spin
moment, $M_t$, follows a Slater-Pauling behavior and it equals in
$\mu_B$ units: $M_t=Z_t-8$ where $Z_t$ the total number of valence
electrons in the unit cell.

As we mentioned above most of the studies concern cases where the
$sp$ atom belongs to the Vth (pncitides) or the VIth
(chalcogenides) column of the periodical table. The case where the
$sp$ atom comes from the IVth column of the periodical table has
attracted much less attention in literature and to the best of our
knowledge only the cases of MnC \cite{Qian,Li08,Pask}, MnSn
\cite{Li08} and MnSi \cite{Wu,Hortamani1,Hortamani2} in the
zincblende lattice have been studied. In this manuscript we study
using the full--potential nonorthogonal local--orbital
minimum--basis band structure scheme (FPLO) \cite{fplo1} within
the local density approximation (LDA) \cite{perdew} the case of
[Sc$_{1-x}$V$_x$]C and [Sc$_{1-x}$V$_x$]Si alloys for $x$ taking
the values 0, 0.1, 0.2, ...,0.9, 1. The disorder is simulated
using the coherent potential approximation \cite{fplo2}. ScC and
ScSi compounds have 7 electrons per unit cell and, if they are
half metals, they should show a total spin moment of -1 $\mu_B$.
On the other hand VC and VSi have 9 electrons and, if they are
half-metals, should exhibit a total spin moment of +1 $\mu_B$.
Thus within these families of alloys we can study the transition
at the $x=0.5$ concentration where the total spin moment changes
sign. In section \ref{sec2} we study the perfect compounds to
determine the equilibrium lattice constants and we show that all
four ScC, ScSi, VC and VSi crystallize in the non-magnetic
rocksalt structure. In section \ref{sec3} we continue our study
with the case of the magnetic zincblende and wurtzite structures
which are degenerated with respect to their total energy and we
study both the electronic and magnetic properties. We show that
for the intermediate concentrations the total spin moment scales
linearly with the concentration and the Sc and V atoms have
antiparallel spin magnetic moments. We also try to explain why for
$x=0.5$ the C-based alloy is a semiconductor while the Si-based
alloys is a half-metallic antiferromagnet \cite{Leuken}. The
latter property is highly desirable for applications since such
materials create vanishing stray fields and thus minimize energy
losses in devices. Finally in section \ref{sec4} we summarize and
present our conclusions.

\begin{figure}
\begin{center}
\includegraphics[scale=0.5]{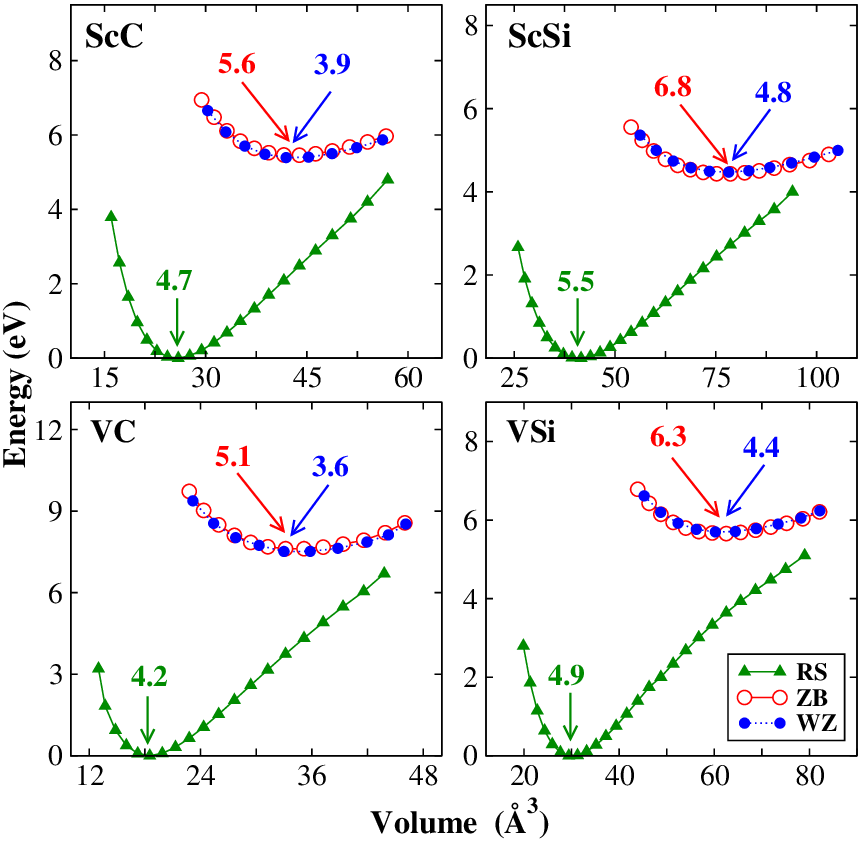}
\end{center} \caption{(Color online)  Calculated total energy as a function of
the volume of the unit cell for ScC, ScSi, VC and VSi in the
rocksalt (RS), zincblende (ZB) and wurtzite (WZ) structures. The
zero of the total energy is defined as the energy of the global
equilibrium volume and with arrows we represent the corresponding
equilibrium lattice constant; for the WZ structure which is not
cubic we give the in-plane lattice parameter $a$ and the $c/a$
ratio is for all calculations the ideal
$(\frac{8}{3})^{\frac{1}{2}}$ for which the nearest environment in
the WZ structure is the same with its cubic ZB analogue. In the RS
and ZB there is one transition-metal atom and one sp atom per unit
cell. In the WZ structure there are two atoms of each chemical
kind but we have divided the energy by two to compare it directly
to the other two cases. \label{fig1}}
\end{figure}

\section{Total energy calculations} \label{sec2}

\begin{figure}
\includegraphics[scale=0.5]{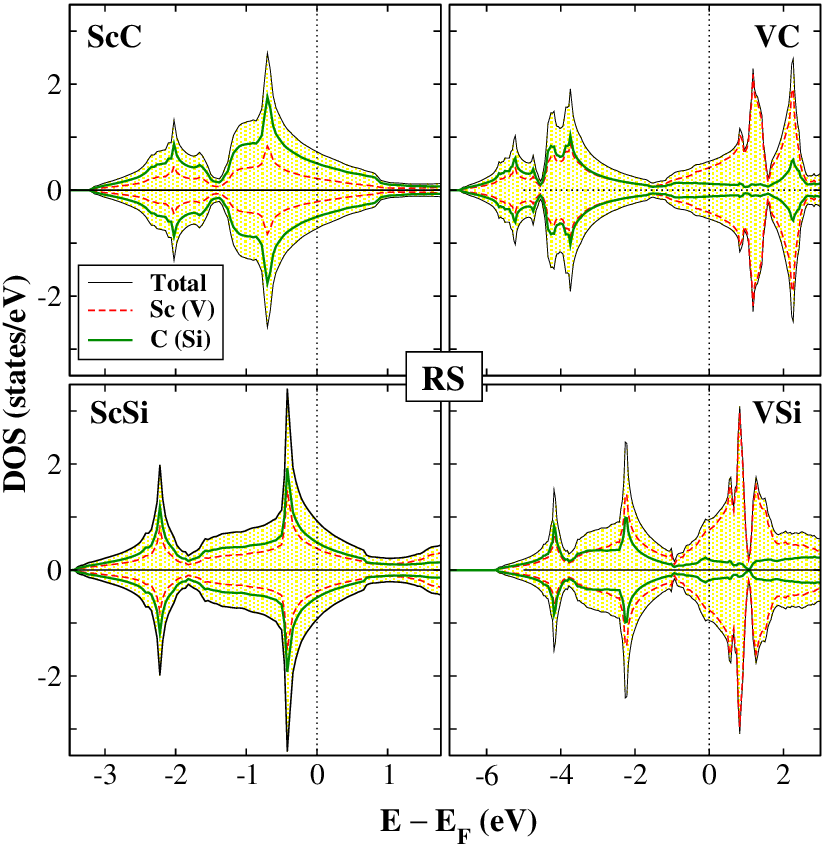}
\caption{(Color online) Total and atom-resolved density of states
(DOS) for all four perfect compounds in the rock-salt structure at
their equilibrium lattice constants. We present upwards the DOS
for the spin-up states and downwards for the spin-down. In this
structure we have converged to a non-magnetic state for all four
alloys. The Fermi level has been chosen as the zero of the energy
axis. \label{fig2}}
\end{figure}

We will start the presentation of our results discussing the
equilibrium lattices and lattice constants. We have taken into
account three different lattices: (i) the rocksalt (RS), (ii) the
zincblende, and (iii) the wurtzite (WZ) structures, and we present
in figure \ref{fig1} the calculated total energy as a function of
the volume of the unit cell. Before proceeding with the discussion
and presentation of our results we should focus on the
characteristics of the three structures under study. We have
chosen the RS structure since it is adopted by the majority of the
binary compounds between transition-metal and $sp$ atoms when the
stoichiometry is 1:1. The lattice is actually a fcc  with two
atoms as basis set, one atom at (0 0 0) and the second atom at
($\frac{1}{2}$ $\frac{1}{2}$ $\frac{1}{2}$) in Wyckoff
coordinates. Thus the RS is a close packed structure. Contrary to
RS both ZB and WZ structures are open structures. In the ZB
structure the lattice is again a fcc with four sites as basis set
along the diagonal, but now two out of the four sites are empty.
The WZ is the hexagonal analogue of the ZB structure. In our WZ
calculations we have varied only the in-plane lattice parameter
$a$ and we have considered that the $c/a$ ratio is for all
calculations the ideal $(\frac{8}{3})^{\frac{1}{2}}$ for which the
nearest environment in the WZ structure is the same with its cubic
ZB analogue. Finally we should mention that in the RS and ZB there
is one transition-metal atom and one sp atom per unit cell while
in the WZ structure there are two atoms of each chemical kind. We
have divided in the WZ case by two all the properties, which are
calculated per unit cell (total density of states, total spin
magnetic moment and total energy),  in order to compare them
directly to the other two cases.

For all four perfect compounds presented in figure \ref{fig1} the
RS structure is the equilibrium lattice and the energy difference
between the equilibrium RS lattice constant and the ZB-WZ
equilibrium lattice constants is between 4 and 8 eV which are very
large energy differences. Thus in the form of single crystals all
ScC, ScSi, VC and VSi prefer to crystallize in the RS lattice. As
expected the equilibrium volume is much smaller in the RS case
compared to the other two since in the former one no voids exist.
Surprisingly our results suggest that the ZB and WZ structure are
degenerated in all cases and thus, when grown as thin films on top
of semiconductors, the unit cell can easily deform itself. In the
same figure we have also denoted the equilibrium lattice constants
for all three structures (for the WZ one we give the in-plane
lattice parameter). We remark that the trends depend on the
chemical elements. When we substitute Si for C, the equilibrium
lattice parameters increases. C and Si have both four valence
electrons but C has six electrons in total (the atomic
configuration is $1s^2$ $2s^2$ $2p^2$) while Si has 14 electrons
(the atomic configuration is $1s^2$ $2s^2$ $2p^6$ $3s^2$ $3p^2$)
and thus occupies more space. Sc and V belong in the same series
in the periodic table. Sc has three valence electrons (the valence
electrons in the atomic configuration are the $4s^2$ and $3d^1$)
and V has two more $d$ electrons and thus in total five valence
electrons. Although V has more valence electrons than Sc, it is
well known that for the early transition metal atoms as the
valence increases the lattice parameter decreases \cite{Moruzzi}
and this is also the cases here.  VC and VSi correspond to smaller
equilibrium lattice parameters than ScC and ScSi, respectively.

Finally we should discuss the electronic properties in the case of
the RS structure before proceeding in the next section with the ZB
and WZ structures. In the RS lattice all four perfect compounds
are non-magnetic as can be seen from the density of states (DOS)
presented in figure \ref{fig2}. In the case of the ScC and ScSi
alloys the Fermi level crosses the valence band, which is created
by the bonding states due to the hybridization between the Sc $d$-
and the C(Si) $p$-orbitals, and the atom-resolved DOS has a
similar shape for both the Sc and C(Si) atoms. When we substitute
V for Sc we populate also partially the conduction band created by
the antibonding $p-d$ states. As can be seen in the figure, the
conduction band has its main weight mainly at the vanadium atom
since the latter one offers a lot of empty $d$-states with respect
to the empty $p$ states of the $sp$ atoms. Finally we should note
that for all four alloys there is one $s$ band lying very deep in
energy and which is separated by more than 3 eV from the bottom of
the valence band and thus we do not present it in figure
\ref{fig2}.

\section{Ferrimagnetism in the zincblende and wurtzite
structures}\label{sec3}

\begin{table}
\centering \caption{Total and atom-resolved spin magnetic moments
in $\mu_B$ for all compounds under study at the equilibrium
lattice constants for the zincblende (ZB) and wurtzite (WZ)
structures. For the intermediate concentrations we have considered
that the lattice constant scales linearly with the concentration.
We have scaled the atom-resolved spin moments to one atom. The
total spin moment is the sum
$(1-x)*m^{Sc}+x*m^{V}+m^{C(Si)}+m^{int}$, where $m^{int}$ refers
to the interstitial region (empty sites). We do not present
$m^{int}$ separately since it is negligible with respect to the
atomic spin moments. In the ZB there is one transition-metal atom
and one sp atom per unit cell, while in the WZ structure there are
two equivalent atoms of each chemical kind and thus we present
half the total spin moment in the unit cell to compare it directly
to the ZB case. }
 \begin{tabular}{l|cccc|cccc} \hline \hline

 & \multicolumn{8}{c}{[Sc$_{1-x}$V$_x$]C } \\

 & \multicolumn{4}{c|}{zincblende} & \multicolumn{4}{c}{wurtzite}
\\

$x$   & $m^{Sc}$ & $m^{V}$ & $m^{C}$ & $m^{Total}$  & $m^{Sc}$ &
$m^{V}$ & $m^{C}$ & $m^{Total}$

\\ \hline
 0    & -0.179 &  -   & -0.821 & -1.000  &  -0.186 &
 - &  -0.814 & -1.000
\\
0.1  & -0.177 & 0.994 & -0.740 & -0.799  & -0.201 & 1.122 & -0.731
& -0.800
\\
0.2  & -0.170 & 0.783 & -0.620 & -0.599 & -0.202 & 0.968 & -0.632
& -0.600
\\
0.3 & -0.144 & 0.533 & -0.458 & -0.399  & -0.183 & 0.764 & -0.501
& -0.400 \\ 0.4& $\sim$-0 & $\sim$0 & $\sim$-0 & $\sim$-0 & -0.126
& 0.466 & 0.310 & -0.200
\\
0.5 & $\sim$-0 & $\sim$0& $\sim$-0 & 0  & $\sim$-0 & $\sim$0&
$\sim$-0 & 0
\\
0.6 & -0.064 & 0.646 & -0.162 & 0.200 & -0.090 & 0.755 & -0.217 &
0.200
\\
0.7  & -0.086 & 0.899 & -0.204 & 0.400  & -0.108 & 0.986 & -0.258
& 0.400
\\
0.8 & -0.098 & 1.047 & -0.218 & 0.600  & -0.117 & 1.122 & -0.274 &
0.600
\\
0.9 & -0.107 & 1.144 & -0.219 & 0.800  & -0.122 & 1.215 & -0.282 &
0.800
\\
1 &   - & 1.192 & -0.192 & 1.000  & - & 1.194 & -0.194 & 1.000

 \\ \hline \\ \hline
 & \multicolumn{8}{c}{[Sc$_{1-x}$V$_x$]Si } \\

 & \multicolumn{4}{c|}{zincblende} & \multicolumn{4}{c}{wurtzite}
\\

$x$  & $m^{Sc}$ & $m^{V}$ & $m^{Si}$ & $m^{Total}$ & $m^{Sc}$ & $m^{V}$ & $m^{Si}$ & $m^{Total}$  \\
\hline

0  & -0.349 & - & -0.651 & -1.000 &  -0.367 & - &-0.633 & -1.000
\\
0.1  & -0.383 & 2.030 & -0.658 & -0.799 &  -0.403 & 1.868 & -0.624
& -0.800
\\
0.2  & -0.423 & 1.914 & -0.644 & -0.599 &  -0.436 & 1.778 & -0.607
& -0.599
\\
0.3 &  -0.451 & 1.764 & -0.613 & -0.399 &  -0.457 & 1.656 & -0.577
& -0.399
\\
0.4 &  -0.448 & 1.558 & -0.554 & -0.199&  -0.456 & 1.504 & -0.527
& -0.199
\\
0.5  & -0.405 & 1.325 & -0.460 & -0.001& -0.429 &
1.352 & -0.461 & 0.001 \\
0.6  & -0.413 & 1.324 & -0.429 & 0.200  & -0.417 & 1.308 & -0.418
& 0.200
\\
0.7 & -0.426 & 1.333 & -0.405 & 0.400  & -0.414 & 1.302 & -0.387 &
0.400
\\
0.8 & -0.439 & 1.337 & -0.381 & 0.600 & -0.413 & 1.305 & -0.361 &
0.600
\\
0.9  & -0.451 & 1.335 & -0.356 & 0.800  & -0.414 & 1.310 & -0.338
& 0.800
\\
1  & - & 1.330 & -0.330 & 1.000 & - & 1.288 & -0.288 & 1.000

\\ \hline
\end{tabular}
\label{table1}
\end{table}

\begin{figure*}
\includegraphics[scale=0.5]{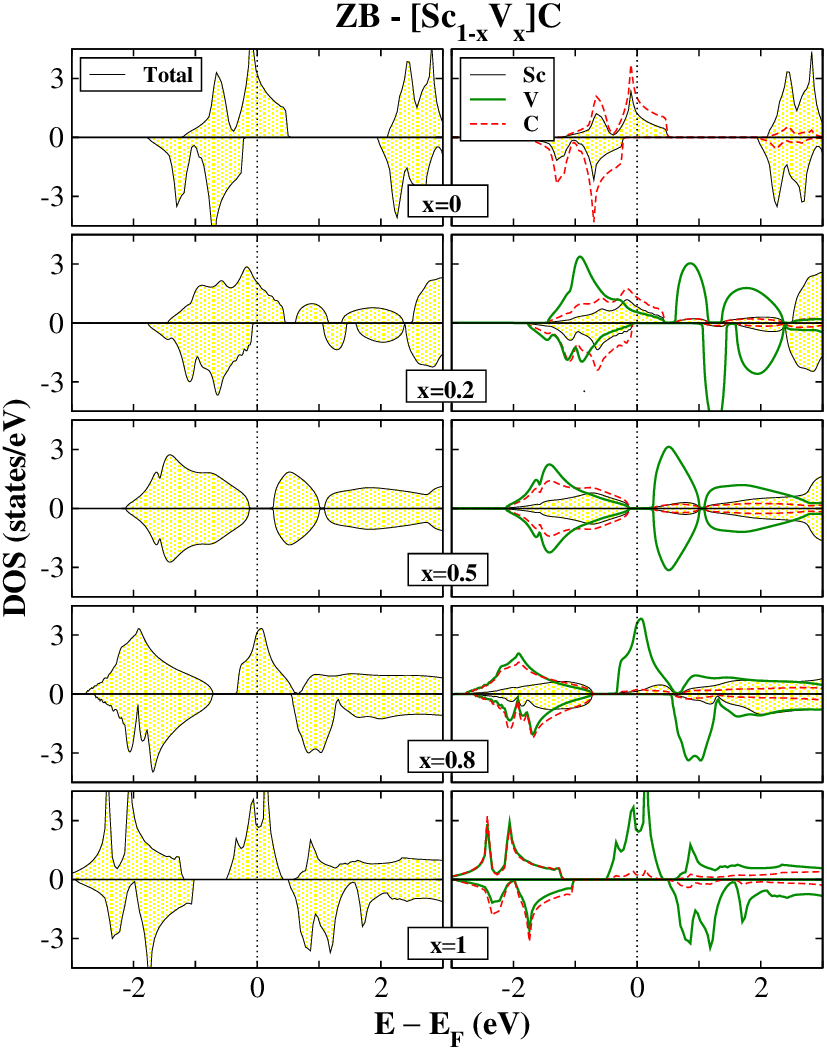} \hskip 0.8cm
\includegraphics[scale=0.5]{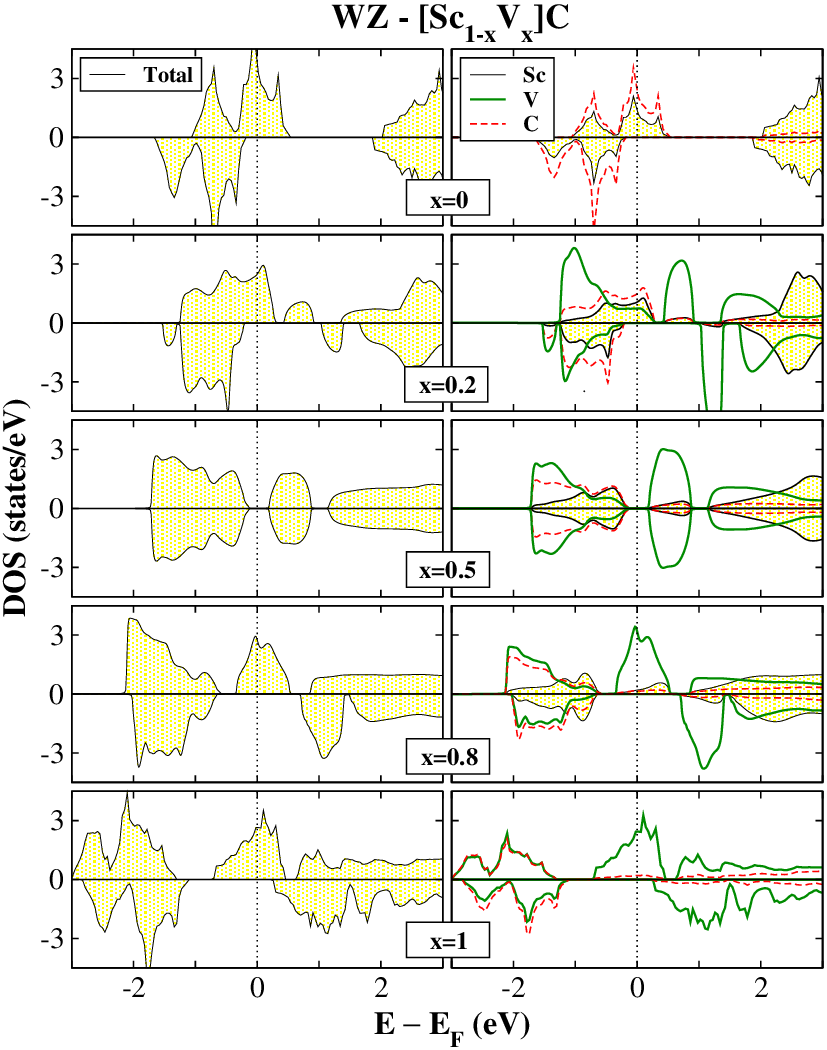}\vskip 0.6cm
\includegraphics[scale=0.5]{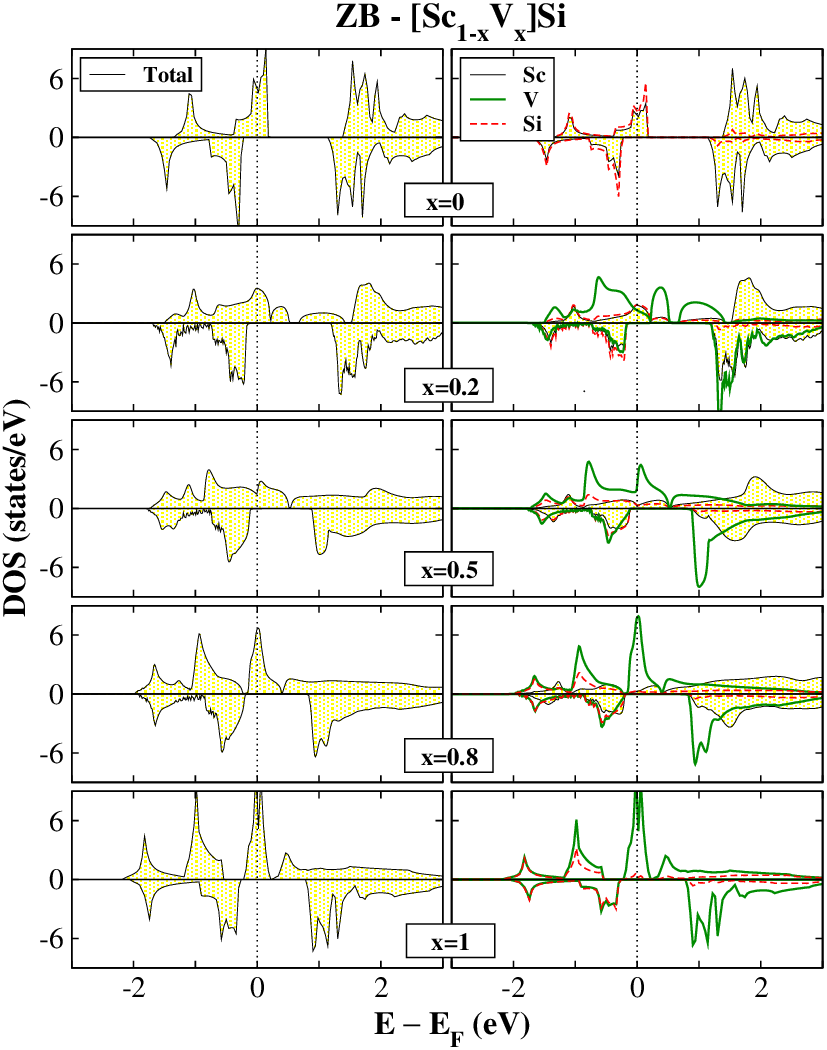} \hskip 0.8cm
\includegraphics[scale=0.5]{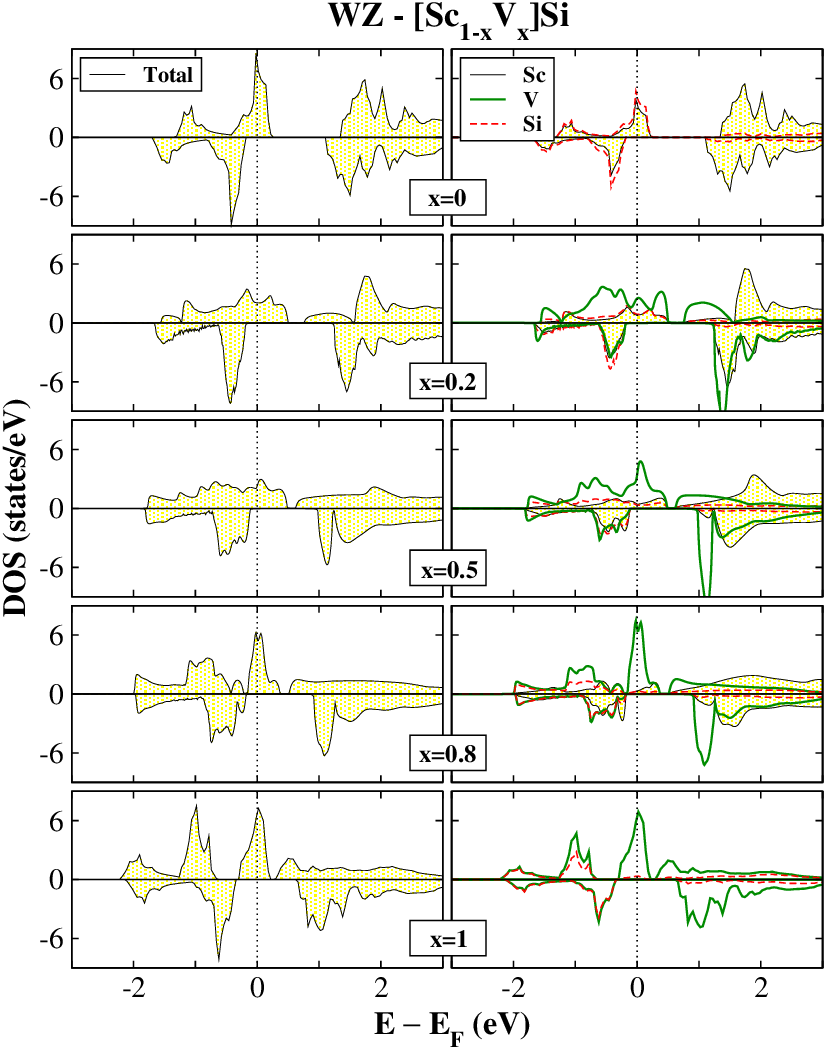}
\caption{(Color online) Total and atom-resolved DOS for
[Sc$_{1-x}$V$_x$]C (upper panel) and [Sc$_{1-x}$V$_x$]Si (lower
panel) in both zincblende (left) and wurtzite (right) lattice
structures for several values of the concentration $x$. The
lattice parameter has been assumed to vary linearly between the
equilibrium lattice constants of the ScC(ScSi) and VC(VSi) alloys.
We have scaled the atom-resolved DOS to one atom. In the case of
the wurtzite structure the unit cell contains double the atoms of
the zincblende unit-cell and thus we have scaled the total DOS in
the wurtzite structure by 0.5 to make them comparable. In all
cases the deep lying $s$ states are not shown. \label{fig3}}
\end{figure*}

As we showed in the previous section all compounds under study are
non-magnetic in the equilibrium RS lattice and thus are not useful
for spintronic applications. Contrary to the RS case, we found all
four ScC, ScSi, VC and VSi to be magnetic in both the ZB and WZ
structures. Although these structures are not stable ones, it is
possible to occur when these alloys are grown as thin films or
multilayers on top of binary semiconductors adopting the same
lattice structure. This is possible with the new up-to-date
experimental techniques like Molecular Beam Epitaxy or Pulsed
Layer Deposition and in fact the former method was employed to
grow the CrAs films with the ZB structure on top of the GaAs
semiconductor in reference \cite{Akinaga}.

We will start our disucssion from the properties of the perfect
compounds and in table \ref{table1} we have gathered the
atom-resolved and total spin magnetic moments and in figure
\ref{fig3} we present the associated DOS. All presented
calculations have been performed at the equilibrium lattice
constants presented in figure \ref{fig1}. Our first remark
concerns the difference between the ZB and WZ structures. As we
can see in table \ref{table1}, the spin magnetic moments only
scarcely vary between the two lattice structures and the same
occurs also for the total and atom-resolved DOS presented in
figure \ref{fig3} (note that we have divided the total spin moment
and DOS in the WZ structure by two to make it comparable to the ZB
case). Thus we will concentrate our discussion to the ZB case and
refer to the WZ structure only if the difference with respect to
the ZB lattice is significant. ScC has in total 7 valence
electrons per unit cell and thus in order to be half-metal it
should exhibit a spin magnetic moment of -1 $\mu_B$ according to
the $M_t=Z_t-8$ Slater-Pauling rule. This is actually the case
since, as can be seen in table \ref{table1}, Sc carries as spin
moment of around -0.18 $\mu_B$ and C atoms carry a spin moment of
around -0.82 $\mu_B$. Although it seems strange that the spin
moment is mainly concencentrated to the $sp$ atoms, this
phenomenon can be easily understood if we look at the DOS in
figure \ref{fig3}. In the spin-down band all four states are
occupied. These states, as we have already discussed in section
\ref{sec1} and as it is shown in reference \cite{GalaZB}, consist
of a deep lying $s$-state and three bonding states due to the
hybridization between the $p$ electrons of the C and the $t_{2g}$
$d$-electrons of the Sc atom. These bonding  states are located
mainly at the C atom since the $p$ states of C lie lower in energy
with respect to the $t_{2g}$ states of Sc. Since in total we have
seven electrons and four are already accomodated in the spin-down
bands, only three have to be accomodated in the spin-up band. Thus
in the spin-up band again only the bonding hybrids with their main
weight at the C atom are partially occupied and thus the spin
moment is mainly concentrated at the $sp$-atom. As can be seen in
the same graph the unoccupied states are mainly of Sc character
since they are consisted of the antibonding $p-t_{2g}$ states and
the localized $e_g$ $d-$states of Sc. When we subsitute V for Sc,
the two extra electrons fill exclusively spin-up states since the
half-metallicity is preserved and the Fermi level is again within
the spin-down gap. Thus now all the boinding spin-up states are
occupied and one electron occupies partially the localized spin-up
$e_g$ states of V (in total these states can occupy two electrons
per spin). As a result the Fermi level falls within a large peak
in the spin-up band and almost divides it in two equal parts. The
spin-up peak of the $e_g$ states is clearly separated by the
antibonding $p-t_{2g}$ states as can be seen in the DOS. As a
result of the above discussion V has now a spin magnetic moment of
$\sim$ 1.19 $\mu_B$ and C a spin magnetic moment of $\sim$ -0.19
$\mu_B$ resulting to a total spin moment of +1 $\mu_B$ in
agreement to the Slater-Pauling behavior. The spin moment of V can
be decomposed to 1 $\mu_B$ due to the $e_g$ state, and 0.19
$\mu_B$ which controbalances the -0.19 $\mu_B$ of C and come from
the small inbalance in the distribution of the bonding $p-t_{2g}$
states between the two chemical species in the spin-up and
spin-down bands. In the case of the Si compound the situation is
similar with only noticeable difference the fact that Si carries
in the case of ScSi a smaller absolute value of the spin magnetic
moment with respect to C in ScC while the situation is inversed
when we compare the VSi to the VC alloy. ScSi and VSi compounds
have significantly larger equilibrium lattice constants with
respect to the ScC and VC alloys as shown in figure \ref{fig1}.
Thus the hybridization between the Si $p$-states and the
$t_{2g}$-states of Sc(V) is less intense than in the case of the
C-based alloys resulting in bands which are more narrow in energy
width but more intense (the scale in the vertical DOS axis in
figure \ref{fig3} is double for the Si-based alloys). Due to this
small change in hybridization a variation in the spin magnetic
moments occurs with respect to the C-based compounds.

We have also performed calculations for the intermediate
concentrations $x$ and in table \ref{table1} and figure \ref{fig3}
we have gathered the atom-resolved and total spin magnetic moments
and DOS. The atom-resolved properties have been scaled to one
atom. We have assumed that the lattice constants scale linearly
with the concentration $x$ between the perfect compounds as occurs
experimentally for the quaternary Heusler alloys which have
similar structure \cite{landolt}. In order for these alloys to be
half-metallic, the Slater-Pauling rule should be again valid where
as valence of the transition-metal site we consider the
$(1-x)*Z_{Sc}+x*Z_V$, where $Z_{Sc}=3$ and $Z_V=5$ the valence of
the Sc and V atoms respectvely. As we dope ScC and ScSi with V,
the bonding $p-t_{2g}$ states in the spin-up band start to move
lower in energy with respect to the Fermi level and the peak of
the $e_g$ states starts appearing at the Fermi level as can be
clearly seen for the case of $x=0.8$. Thus almost all intermediate
compounds follow the Slater-Pauling behavior as can be seen from
the total spin moments in table \ref{table1} and the electronic
and magnetic properties of the compounds vary in a continuous way
with the concentration. The compounds are ferrimagnets since the V
atoms have a spin moment antiparallel to the one of the Sc and
C(Si) atoms.

The only exception of compounds which are not magnetic is the case
of [Sc$_{1-x}$V$_x$]C for $x=$0.4 and 0.5 in the ZB lattice and
$x=$0.5 in the WZ structure. We will start our  discussion from
the case of $x=$0.5. [Sc$_{0.5}$V$_{0.5}$]C is not magnetic both
in the ZB and WZ structures and as shown in the DOS presented in
figure \ref{fig3} it is actually a semiconductor since there are
exactly eight valence electrons which occupy the bonding
$p-t_{2g}$ states in both the spin-up and spin-down bands. The
energy gap is about 0.4 eV. We can clearly distinguish in the DOS
the occupied bonding hybrids, followed by the $e_g$ bands just
above the Fermi level with most of their weight at the V atoms and
which are clearly separated in energy from the antibonding
$p-t_{2g}$ states which are even higher in energy (the deep-lying
$s$ states are not shown). In the case of $x=$0.4 in the ZB
structure the Fermi level falls in a region of small spin-up DOS
(the DOS is not presented here) and due to the Stoner theorem the
alloy is not magnetic contrary to the WZ structure where the
Stoner theorem is satisfied even marginally resulting in a
magnetic compound.

\begin{figure}
\includegraphics[scale=0.5]{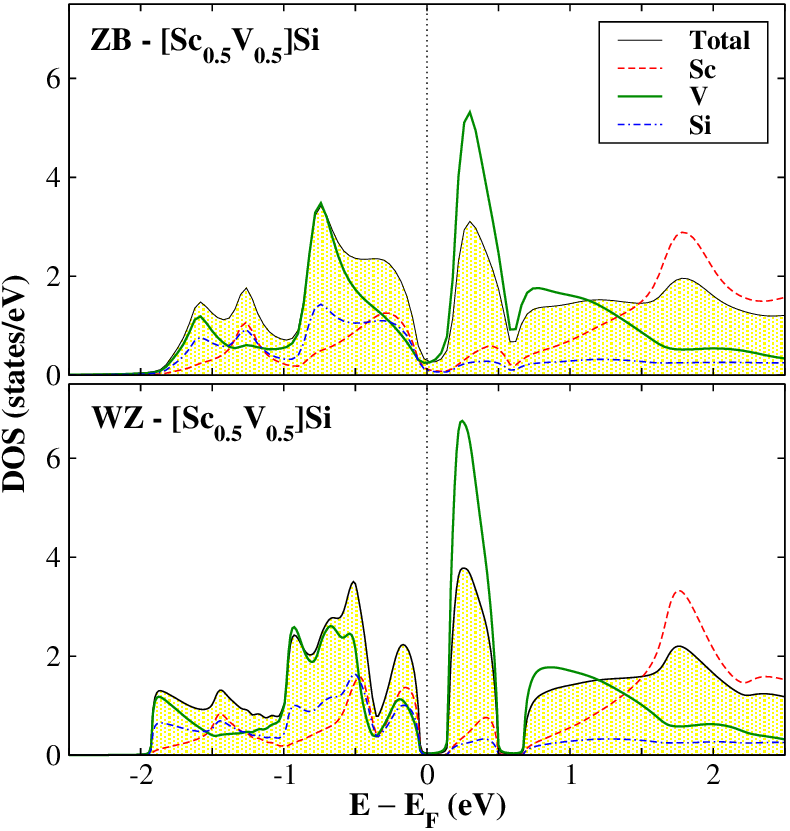}
\caption{(Color online) Total and atom-resolved DOS for
[Sc$_{0.5}$V$_{0.5}$]Si in both zincblende and wurtzite structures
performing non-spinpolarized calculations. The atom-resolved DOS
have been scaled to one atom. \label{fig4}}
\end{figure}

Contrary to [Sc$_{0.5}$V$_{0.5}$]C the [Sc$_{0.5}$V$_{0.5}$]Si
alloy is magnetic and in reallity it is a so-called half-metallic
antiferromagnet \cite{Leuken,GalaAFM} since the total spin-moment
is exactly zero as in conventional antiferromagnets but the
different constituents are magnetic as shown in table
\ref{table1}. The question which arises is why the Si-alloy shows
such a different behavior from the C-alloy. To elucidate the
origin of the difference we have performed non-magnetic
calculations for the [Sc$_{0.5}$V$_{0.5}$]Si alloy in both the ZB
and WZ structures and we present the DOS in figure \ref{fig4}. The
Fermi level falls within a deep of the DOS in the ZB structure and
within a small gap in the WZ structure, thus due to Stoner theorem
in both cases magnetism is not favorable. But Stoner theorem has a
limited application in alloys where the situation is more
complicated with respect to crystals made out of a single chemical
element. Below the Fermi level the  bands are concentrated in a
small energy range and just above the Fermi level there is a very
intense peak due to the V $e_g$ states making the non-magnetic
solution unstable and the system prefers to be magnetic in order
to lower its total energy.

\section{Summary and conclusions}\label{sec4}

We have employed first-principles calculations and have studied
the structural, electronic and magnetic properties of the
[Sc$_{1-x}$V$_x$]C and [Sc$_{1-x}$V$_x$]Si alloys. In their
equilibrium rocksalt structure all alloys are non-magnetic. The
zincblende and wurtzite structures are degenerated with respect to
the total energy and show magnetism. For all concentrations we
found that the alloys in these two lattice structures are
half-metallic with the gap located in the spin-down band. Moreover
they are ferrimagnets since V and Sc atoms have antiparallel
spin-magnetic moments. The total spin moment follows the
Slater-Pauling behavior varying linearly between the -1 $\mu_B$ of
the perfect ScC and ScSi alloys and the +1 $\mu_B$ of the perfect
VC and VSi alloys.  At the critical concentration, both
[Sc$_{0.5}$V$_{0.5}$]C and [Sc$_{0.5}$V$_{0.5}$]Si alloys present
zero total spin-magnetic moment but the C-based alloy shows a
semiconducting behavior since the Stoner criterion for magnetism
is not satisfied. Contrary, the [Sc$_{0.5}$V$_{0.5}$]Si compounds
is a half-metallic antiferromagnet and thus very interesting for
potential application in spintronics.

We have shown that also in the case of artificial binary compounds
between transition-metal and $sp$ atoms we can tune their magnetic
properties in a continuous way by mixing atoms of neighboring
chemical species. Following this procedure we were able to
demonstrate the existence of half-metallic ferrimagnetism in the
studied alloys (and in an extreme case even half-metallic
antiferromagnetism) which is highly desirable for spintronic
applications since ferrimagnets create smaller stray fields than
ferromagnets and consequatively lead to smaller energy losses in
devices.

 \ack{Authors
acknowledge the computer support of the Leibniz Institute for
Solid State and Materials Research Dresden, and the assistance of
Ulrike Nitzsche in using the computer facilities.}


\end{document}